\documentclass[runningheads]{llncs}
\usepackage{graphicx}
\usepackage[misc]{ifsym}
\usepackage{amsmath}
\usepackage{amssymb}
\usepackage{multirow}
\usepackage{bbding}
\usepackage{pifont}
\usepackage{wasysym}
\usepackage{utfsym}
\usepackage{fontawesome}
\usepackage{url}
\usepackage{subcaption}
\usepackage{mathtools}
\usepackage{color}
\usepackage{booktabs}
\usepackage{makecell,rotating}
\usepackage[misc]{ifsym}
\usepackage{threeparttable}

\def\etal{\textit{et al}. }

\def\eg{\textit{e.g. }}

\usepackage{hyperref}
\hypersetup{hidelinks,
	colorlinks=true,
	allcolors=black,
	pdfstartview=Fit,
	breaklinks=true}

\begin{document}
%

\title{DHC: Dual-debiased Heterogeneous Co-training Framework for Class-imbalanced Semi-supervised Medical Image Segmentation}

\titlerunning{Dual-debiased Heterogeneous Co-training Framework}
%

\authorrunning{H. Wang \& X. Li}

\author{Haonan Wang, Xiaomeng Li\textsuperscript{(\Letter)}} 

\institute{Department of Electronic and Computer Engineering, The Hong Kong University
of Science and Technology, Hong Kong, China
 \\ \email{eexmli@ust.hk}}


%
\maketitle              

\begin{abstract}
The volume-wise labeling of 3D medical images is expertise-demanded and time-consuming; hence semi-supervised learning (SSL) is highly desirable for training with limited labeled data. 
\textit{Imbalanced class distribution} is a severe problem that bottlenecks the real-world application of these methods but was not addressed much.
Aiming to solve this issue, we present a novel \textbf{D}ual-debiased \textbf{H}eterogeneous \textbf{C}o-training (\textbf{DHC}) framework for semi-supervised 3D medical image segmentation. 
Specifically, we propose two loss weighting strategies, namely Distribution-aware Debiased Weighting (DistDW) and Difficulty-aware Debiased Weighting (DiffDW), which leverage the pseudo labels dynamically to guide the model to solve data and learning biases.
The framework improves significantly by co-training these two \textit{diverse and accurate} sub-models.
We also introduce more representative benchmarks for class-imbalanced semi-supervised medical image segmentation, which can fully demonstrate the efficacy of the class-imbalance designs.
Experiments show that our proposed framework brings significant improvements by using pseudo labels for debiasing and alleviating the class imbalance problem. 
More importantly, our method outperforms the state-of-the-art SSL methods, demonstrating the potential of our framework for the more challenging SSL setting. Code and models are available at: \href{https://github.com/xmed-lab/DHC}{https://github.com/xmed-lab/DHC}
\end{abstract}

\section{Introduction} \label{intro}
\begin{figure}[t]
\centering
\includegraphics[width=\linewidth]{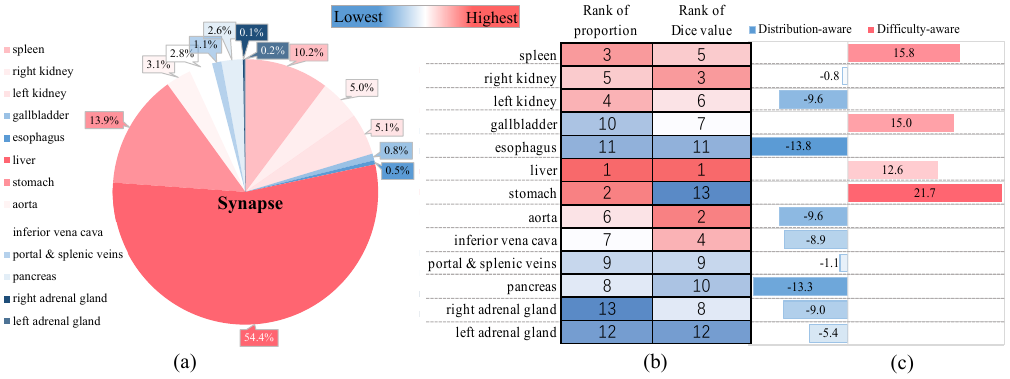} 
\caption{(a) Foreground classes distributions in Synapse dataset.  (b) Comparison of the ranks of proportions and Dice value of proposed DistDW of each class. (c) Comparison between number of voxels and Dice value of each class, values of x-axis are the difference of Dice values between DiffDW and DistDW models.}
\label{analysis}
\end{figure}

The shortage of labeled data is a significant challenge in medical image segmentation, as acquiring large amounts of labeled data is expensive and requires specialized knowledge. This shortage limits the performance of existing segmentation models. 
To address this issue, researchers have proposed various semi-supervised learning (SSL) techniques that incorporate both labeled and unlabeled data to train models for both natural~\cite{tarvainen2017meanteacher,sohn2020fixmatch,chen2021cps,wang2022u2pl,wang2022depl,chen2022dst} and medical images~\cite{yu2019uamt,luo2021urpc,wang2022gbdl,wu2022cdcl,lin2022cld,you2022cvrl,wu2022ssnet}. 
However, most of these methods do not consider the class imbalance issue, which is common in medical image datasets. For example, multi-organ segmentation from CT scans requires to segment esophagus, right adrenal gland, left adrenal gland, \textit{etc.}, where the class ratio is quite imbalanced; see Fig~\ref{analysis}(a). As for liver tumor segmentation from CT scans, usually the ratio for liver and tumor is larger than 16:1.

\if 1
Recently, more attention has been given to addressing the class imbalance problem in semi-supervised classification tasks through techniques such as leveraging unlabeled data~\cite{wei2021crest,fan2022cossl,simis}, re-balancing data distributions in loss~\cite{hong2021lade}, and re-sampling learning~\cite{yu2022instancediff}. However, directly adopting these methods to medical image segmentation leads to unsatisfactory results (Table~\ref{sota}, especially those colored with \textcolor{red}{red}), which are mainly due to the more imbalanced classes and complex features in medical images. 
Thus, designing a robust class-imbalanced semi-supervised segmentation framework tailored for medical images is necessary.
\fi 


Recently, some researchers proposed class-imbalanced semi-supervised methods~\cite{basak2022addressing,lin2022cld} and demonstrated substantial advances in medical image segmentation tasks. 
Concretely, Basak \etal~\cite{basak2022addressing} introduced a robust class-wise sampling strategy to address the \textit{learning bias} by maintaining performance indicators on the fly and using fuzzy fusion to dynamically obtain the class-wise sampling rates. 
However, the proposed indicators can not model the difficulty well, and the benefits may be overestimated due to the non-representative datasets used (Fig.~\ref{analysis}(a)).
Lin \textit{et al.}~\cite{lin2022cld} proposed CLD to address the \textit{data bias} by weighting the overall loss function based on the voxel number of each class. However, this method fails due to the easily over-fitted CPS (Cross Pseudo Supervision)~\cite{chen2021cps} baseline, ignoring unlabeled data in weight estimation and the fixed class-aware weights. 


\begin{figure}[t]
\centering
\includegraphics[width=0.9\linewidth]{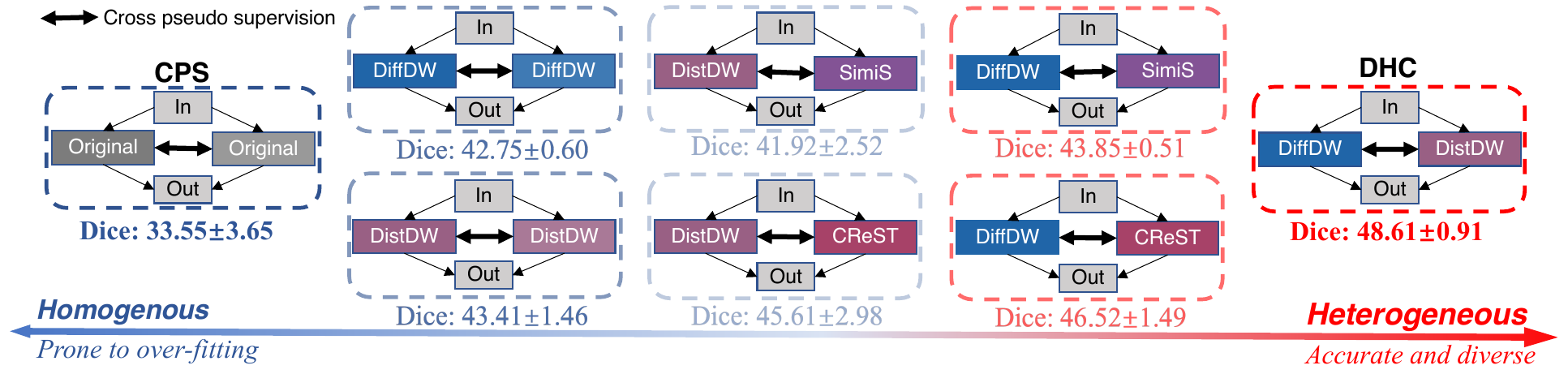} 
\caption{Illustration of the homogeneity problem of CPS and the effectiveness of using heterogeneous framework.}
\label{heterogeneous}
\end{figure}

In this work, we explore the importance of heterogeneity in solving the over-fitting problem of CPS (Fig.~\ref{heterogeneous}) and propose a novel \textbf{DHC} (\textbf{D}ual-debiased \textbf{H}eterogeneous \textbf{C}o-training) framework with two distinct dynamic weighting strategies leveraging both labeled and unlabeled data, to tackle the class imbalance issues and drawbacks of the CPS baseline model.
The key idea of heterogeneous co-training is that individual learners in an ensemble model should be both \textit{accurate and diverse}, as stated in the error-ambiguity decomposition~\cite{krogh1994neural}. 
To achieve this, we propose \textbf{DistDW} (\textbf{Dist}ribution-aware \textbf{D}ebiased \textbf{W}eighting) and \textbf{DiffDW} (\textbf{Diff}iculty-aware \textbf{D}ebiased \textbf{W}eighting) strategies
to guide the two sub-models to tackle different biases, leading to heterogeneous learning directions.
Specifically, DistDW solves the \textit{data bias} by calculating the imbalance ratio with the unlabeled data and forcing the model to focus on extreme minority classes through careful function design.
Then, after observing the inconsistency between the imbalance degrees and the performances (see Fig.~\ref{analysis}(b)), DiffDW is designed to solve the \textit{learning bias}. We use the labeled samples and the corresponding labels to measure the learning difficulty from learning speed and Dice value aspects and slow down the speeds of the easier classes by setting smaller weights. 
DistDW and DiffDW are diverse and have complementary properties (Fig.~\ref{analysis}(c)), which satisfies the design ethos of a heterogeneous framework.

The key contributions of our work can be summarized as follows: 1) we first state the homogeneity issue of CPS and improve it with a novel dual-debiased heterogeneous co-training framework targeting the class imbalance issue; 2) we propose two novel weighting strategies, DistDW and DiffDW, which effectively solve two critical issues of SSL: data and learning biases; 3) we introduce two public datasets, Synapse~\cite{synapse} and AMOS~\cite{amos}, as new benchmarks for class-imbalanced semi-supervised medical image segmentation.
These datasets include sufficient classes and significant imbalance ratios ($>500:1$), making them ideal for evaluating the effectiveness of class-imbalance-targeted algorithm designs.

\section{Methods}

\begin{figure}[t]
\centering
\includegraphics[width=\linewidth]{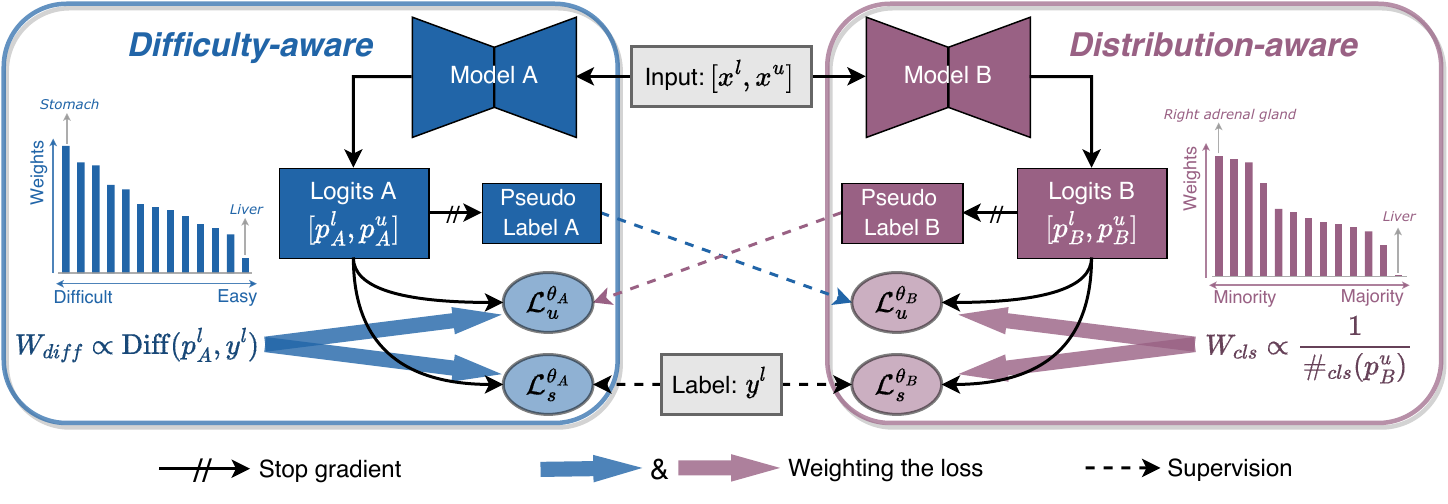} 
\caption{Overview of the proposed dual-debiased heterogeneous co-training framework.}
\label{framework}
\end{figure}

Fig.~\ref{framework} shows the overall framework of the proposed DHC framework. DHC leverages the benefits of combining two \textit{diverse and accurate} sub-models with two distinct learning objectives: alleviating data bias and learning bias.
To achieve this, we propose two dynamic loss weighting strategies, DistDW (Distribution-aware Debiased Weighting) and DiffDW (Difficulty-aware Debiased Weighting), to guide the training of the two sub-models.
DistDW and DiffDW demonstrate complementary properties.
Thus, by incorporating multiple perspectives and sources of information with DistDW and DiffDW, the overall framework reduces over-fitting and enhances the generalization capability.

\subsection{Heterogeneous Co-training Framework with Consistency Supervision}



Assume that the whole dataset consists of $N_L$ labeled samples $\{(x_i^l,y_i)\}_{i=1}^{N_L}$ and $N_U$ unlabeled samples $\{x_i^u\}_{i=1}^{N_U}$, where $x_i \in \mathbb{R}^{D\times H\times W}$ is the input volume and $y_i \in \mathbb{R}^{K\times D\times H\times W}$ is the ground-truth annotation with $K$ classes (including background). 
The two sub-models of DHC complement each other by minimizing the following objective functions with two \textit{diverse and accurate} weighting strategies:
\begin{equation}
    \overline{\mathcal{L}_s}=\frac{1}{N_L}\frac{1}{K}\sum_{i=0}^{N_L} [ {\color[RGB]{33,101,168} W^{diff}_i}\mathcal{L}_{s}(p^A_i, y_i)+ { \color[RGB]{153,97,132}W^{dist}_i}\mathcal{L}_{s}(p^B_i, {y}_i)]
 \end{equation}
\begin{equation}
    \overline{\mathcal{L}_u}=\frac{1}{N_L+N_U}\frac{1}{K}\sum_{i=0}^{N_L+N_U} [{\color[RGB]{33,101,168}W^{diff}_i}\mathcal{L}_{u}(p^A_i, \hat{y}_i^B)+ { \color[RGB]{153,97,132}W^{dist}_i}\mathcal{L}_{u}(p^B_i, \hat{y}_i^A)]
\end{equation}
where $p_i^{(\cdot)}$ is the output probability map and $\hat{y}_i^{(\cdot)} = \mathbf{argmax}\{p_{i,k}^{(\cdot)}\}_{k=0}^K$ is the pseudo label of $i^{th}$ sample.
$\mathcal{L}_s(x,y)=\mathcal{L}_{CE}(x,y)$ is the supervised cross entropy loss function to supervise the output of labeled data, and $\mathcal{L}_u(x,y) = \frac{1}{2}[\mathcal{L}_{CE}(x,y)+\mathcal{L}_{Dice}(x,y)]$ is the unsupervised loss function to measure the prediction consistency of two models by taking the same input volume $x_i$. 
Note that both labeled and unlabeled data are used to compute the unsupervised loss. 
Finally, we can obtain the total loss: $\mathcal{L}_{total}=\overline{\mathcal{L}_s}+\lambda \overline{\mathcal{L}_u}$, we empirically set $\lambda$ as 0.1 and follow~\cite{lin2022cld} to use the epoch-dependent Gaussian ramp-up strategy to gradually enlarge the ratio of unsupervised loss.
${\color[RGB]{33,101,168}W^{diff}_i}$ and ${ \color[RGB]{153,97,132}W^{dist}_i}$ are the dynamic class-wise loss weights obtained by the proposed weighting strategies, which will be introduced next.

\begin{figure}[t]
\centering
\includegraphics[width=\linewidth]{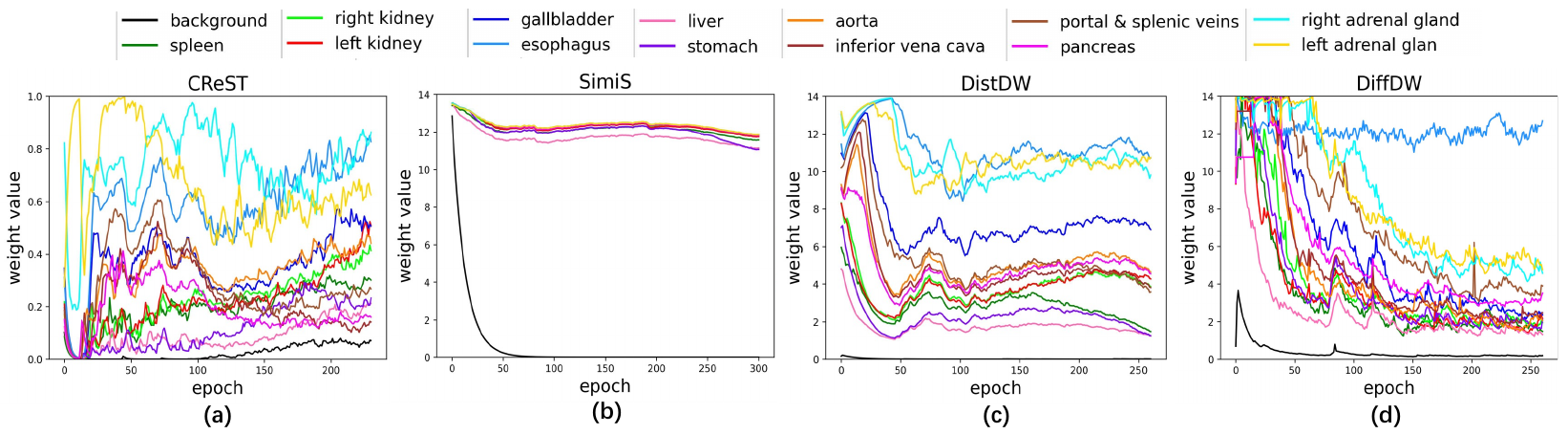} 
\caption{The weight curves of four weighting strategies. SimiS(b) solved the over-weighting issue of the largest majority class (black curve) in CReST(a) but resulted in similar weights for other classes.
DistDW solves the above issues. 
Besides, DiffDW assigns higher weights to difficult classes, such as the \textcolor[RGB]{128,0,255}{stomach}, which has minimum weights when using other methods.}
\label{curves}
\end{figure}

\subsection{Distribution-aware Debiased Weighting (DistDW)}

To mitigate the data distribution bias, we propose a simple yet efficient re-weighing strategy, DistDW. DistDW combines the benefits of the SimiS~\cite{simis}, which eliminate the weight of the largest majority class, while preserving the distinctive weights of the minority classes (Fig.~\ref{curves}(c)). The proposed strategy re-balances the learning process by forcing the model to focus more on the minority classes.
Specifically, we utilize the class-wise distribution of the unlabeled pseudo labels $p^u$ by counting the number of voxels for each category, denoted as $N_k , k=0, . . . , K$. We construct the weighting coeﬃcient for $k^{th}$ category as follows:
\begin{equation}
    w_k= \frac{log(P_k)} {\mathbf{max}\{log(P_i)\}^K_{i=0}}, \quad  P_k = \frac {\mathbf{max}\{N_i\}^K_{i=0}} {N_k}, \quad k=0,1,...,K
\end{equation}
\begin{equation}
    W_t^{dist} \leftarrow \beta W^{dist}_{t-1} + (1-\beta) W^{dist}_t,\quad W_t^{dist} = [w_1, w_2, ..., w_K]
\end{equation}
where $\beta$ is the momentum parameter, set to 0.99 experimentally. 

\subsection{Difficulty-aware Debiased Weighting (DiffDW)}

After analyzing the proposed DistDW, we found that some classes with many samples present significant learning difficulties. For instance, despite having the second highest number of voxels, the stomach class has a much lower Dice score than the aorta class, which has only 20\% of the voxels of the stomach (Fig.~\ref{analysis}(b)). Blindly forcing the model to prioritize minority classes may further exacerbate the learning bias, as some challenging classes may not be learned to an adequate extent.
To alleviate this problem, we design DiffDW to force the model to focus on the most difficult classes (\textit{i.e.} the classes learned slower and with worse performances) rather than the minority classes.
The difficulty is modeled in two ways: learning speed and performance.
We use Population Stability Index~\cite{jeffreys1946psi} to measure the learning speed of each class after the $t^{th}$ iteration:
\begin{equation}
    du_{k,t}=\sum^t_{t-\tau} \mathbb{I}(\triangle \leq 0) \mathrm{ln}(\frac {\lambda_{k,t}} {\lambda_{k,t-1}}), \quad 
    dl_{k,t}=\sum^t_{t-\tau} \mathbb{I}(\triangle > 0) \mathrm{ln}(\frac {\lambda_{k,t}} {\lambda_{k,t-1}})
\end{equation}
where $\lambda_k$ denotes the Dice score of $k^{th}$ class in $t^{th}$ iteration and $\triangle=\lambda_{k,t}-\lambda_{k,t-1}$. $du_{k,t}$ and $dl_{k,t}$ denote classes not learned and learned after the $t^{th}$ iteration. $\mathbb{I(\cdot)}$ is the indicator function. $\tau$ is the number accumulation iterations and set to 50 empirically.
Then, we define the difficulty of $k^{th}$ class after $t^{th}$ iteration as $d_{k,t}= \frac {{du_{k,t}}+\epsilon} {{dl_{k,t}}+\epsilon}$, where $\epsilon$ is a smoothing item with minimal value. The classes learned faster have smaller $d_{k,t}$, the corresponding weights in the loss function will be smaller to slow down the learn speed.
After several iterations, the training process will be stable, and the difficulties of all classes defined above will be similar. Thus, we also accumulate $1-\lambda_{k,t}$ for $\tau$ iterations to obtain the reversed Dice weight $w_{\lambda_{k,t}}$ and weight $d_{k,t}$. 
In this case, classes with lower Dice scores will have larger weights in the loss function, which forces the model to pay more attention to these classes.
The overall difficulty-aware weight of $k^{th}$ class is defined as:
$w_k^{diff}= w_{\lambda_{k,t}} \cdot (d_{k,t})^\alpha$. $\alpha$ is empirically set to $\frac 1 5$ in the experiments to alleviate outliers.
The difficulty-aware weights for all classes are $W_t^{diff} = [w_1, w_2, ..., w_K]$.

\section{Experiments}

\subsubsection{Dataset and Implementation Details.} We introduce two new benchmarks on the Synapse~\cite{synapse} and AMOS~\cite{amos} datasets for class-imbalanced semi-supervised medical image segmentation. 
The Synapse dataset has 13 foreground classes, including spleen (Sp), right kidney ({RK}), left kidney ({LK}), gallbladder ({Ga}), esophagus ({Es}), liver({Li}), stomach({St}), aorta ({Ao}), inferior vena cava ({IVC}), portal \& splenic veins ({PSV}), pancreas ({Pa}), right adrenal gland ({RAG}), left adrenal gland ({LAG}) with one background and 30 axial contrast-enhanced abdominal CT scans.
We randomly split them as 20, 4 and 6 scans for training, validation, and testing, respectively.
Compared with Synapse, the AMOS dataset excludes PSV but adds three new classes: duodenum(Du), bladder(Bl) and prostate/uterus(P/U). 360 scans are divided into 216, 24 and 120 scans for training, validation, and testing.
We ran experiments on Synapse three times with different seeds to eliminate the effect of randomness due to the limited samples.
\textbf{More training details are in the supplementary material.}

\begin{table}[!ht]
\scriptsize
\caption{Quantitative comparison between DHC and SOTA SSL segmentation methods on \textbf{20\% labeled Synapse dataset}. `General' or `Imbalance' indicates whether the methods consider the class imbalance issue or not. Results of 3-times repeated experiments are reported in the `mean$\pm$std' format. Best results are boldfaced, and $2^{nd}$ best results are underlined. 
}
\label{sota1}
\resizebox*{\linewidth}{!}{

\begin{tabular}{c|c|cc|cccccccc@{\ }ccccc}
\toprule
\multicolumn{2}{c|}{\multirow{2}{*}{Methods}}  & \multirow{2}{*}{Avg. Dice} & \multirow{2}{*}{Avg. ASD} & \multicolumn{13}{c}{Average Dice of Each   Class}                                        \\ 
\multicolumn{2}{c|}{}                          &                            &                           & Sp   & RK   & LK   & Ga   & Es   & Li   & St   & Ao   & IVC  & PSV  & PA   & RAG  & LAG  \\\midrule

\multirow{9}{*}{\rotatebox{90}{General}}         & V-Net (fully)      & 62.09±1.2	&10.28±3.9	& 84.6	& 77.2	& 73.8	& 73.3	& 38.2	& 94.6	& 68.4	& 72.1	& 71.2	& 58.2	& 48.5	& 17.9	& 29.0 \\ \midrule

& UA-MT~\cite{yu2019uamt}$^\dagger$      & 20.26±2.2	&71.67±7.4	& 48.2	& 31.7	& 22.2	& \textcolor{red}{0.0}	& \textcolor{red}{0.0}	& 81.2	& 29.1	& 23.3	& 27.5	& \textcolor{red}{0.0}	& \textcolor{red}{0.0}	& \textcolor{red}{0.0}	& \textcolor{red}{0.0}  \\
                 
& URPC~\cite{luo2021urpc}$^\dagger$       & 25.68±5.1	&72.74±15.5	& \textbf{66.7}	& 38.2	& 56.8	& \textcolor{red}{0.0}	& \textcolor{red}{0.0}	& 85.3	& 33.9	& 33.1	& 14.8	& \textcolor{red}{0.0}	& 5.1	& \textcolor{red}{0.0}	& \textcolor{red}{0.0}  \\

& CPS~\cite{chen2021cps}$^\dagger$        & 33.55±3.7	&41.21±9.1	& 62.8	& 55.2	& 45.4	& 35.9	& \textcolor{red}{0.0}	& \textbf{91.1}	& 31.3	& 41.9	& 49.2	& 8.8	& 14.5	& \textcolor{red}{0.0}	& \textcolor{red}{0.0}  \\

& SS-Net~\cite{wu2022ssnet}$^\dagger$    & 35.08±2.8	&50.81±6.5	& 62.7	& 67.9	& \textbf{60.9}	& 34.3	& \textcolor{red}{0.0}	& 89.9	& 20.9	& 61.7	& 44.8	& \textcolor{red}{0.0}	& 8.7	& 4.2	& \textcolor{red}{0.0}  \\

& DST~\cite{chen2022dst}$^\star$        & 34.47±1.6	&37.69±2.9	& 57.7	& 57.2	& 46.4	& 43.7	& \textcolor{red}{0.0}	& 89.0	& 33.9	& 43.3	& 46.9	& 9.0	& \underline{21.0}	& \textcolor{red}{0.0}	& \textcolor{red}{0.0}  \\

& DePL~\cite{wang2022depl}$^\star$       & 36.27±0.9	&36.02±0.8	& 62.8	& 61.0	& 48.2	& 54.8	& \textcolor{red}{0.0}	& \underline{90.2}	& \underline{36.0}	& 42.5	& 48.2	& 10.7	& 17.0	& \textcolor{red}{0.0}	& \textcolor{red}{0.0}  \\ \midrule

\multirow{6}{*}{\rotatebox{90}{Imbalance}} 
& Adsh~\cite{guo2022adsh}$^\star$       & 35.29±0.5	&39.61±4.6	& 55.1	& 59.6	& 45.8	& 52.2	& \textcolor{red}{0.0}	& 89.4	& 32.8	& 47.6	& 53.0	& 8.9	& 14.4	& \textcolor{red}{0.0}	& \textcolor{red}{0.0}  \\ 

& CReST~\cite{wei2021crest}$^\star$      & 38.33±3.4	&\underline{22.85±9.0}	& 62.1	& 64.7	& 53.8	& 43.8	& \underline{8.1}	& 85.9	& 27.2	& 54.4	& 47.7	& 14.4	& 13.0	& \underline{18.7}	& 4.6  \\

& SimiS~\cite{simis}$^\star$      & 40.07±0.6	&32.98±0.5	& 62.3	& \underline{69.4}	& 50.7	& 61.4	& \textcolor{red}{0.0}	& 87.0	& 33.0	& 59.0	& \underline{57.2}	& \underline{29.2}	& 11.8	& \textcolor{red}{0.0}	& \textcolor{red}{0.0}  \\

& Basak \textit{et al.}~\cite{basak2022addressing}$^\dagger$        &33.24±0.6	&43.78±2.5	& 57.4	& 53.8	& 48.5	& 46.9	& \textcolor{red}{0.0}	& 87.8	& 28.7	& 42.3	& 45.4	& 6.3	& 15.0	& \textcolor{red}{0.0}	& \textcolor{red}{0.0}   \\
                                 
& CLD~\cite{lin2022cld}$^\dagger$        &\underline{41.07±1.2}	&32.15±3.3	& 62.0	& 66.0	& \underline{59.3}	& \underline{61.5}	& \textcolor{red}{0.0}	& 89.0	& 31.7	& \underline{62.8}	& 49.4	& 28.6	& 18.5	& \textcolor{red}{0.0}	& \underline{5.0}  \\
 
& \textbf{DHC (ours)}   & \textbf{48.61±0.9}	&\textbf{10.71±2.6}	& \underline{62.8}	& \textbf{69.5}	& 59.2	& \textbf{66.0}	& \textbf{13.2}	& 85.2	& \textbf{36.9}	& \textbf{67.9}	& \textbf{61.5}	& \textbf{37.0}	& \textbf{30.9}	& \textbf{31.4}	& \textbf{10.6} \\ \bottomrule
\end{tabular}
}
\begin{threeparttable}
 \begin{tablenotes}
        \scriptsize
        \item[$\dagger$] we implement semi-supervised segmentation methods on our dataset.
        \item[$\star$] we extend semi-supervised classification methods to segmentation with CPS as the baseline.
\end{tablenotes}
\end{threeparttable}

\end{table}

\begin{figure}[!ht]
\centering
\includegraphics[width=0.95\linewidth]{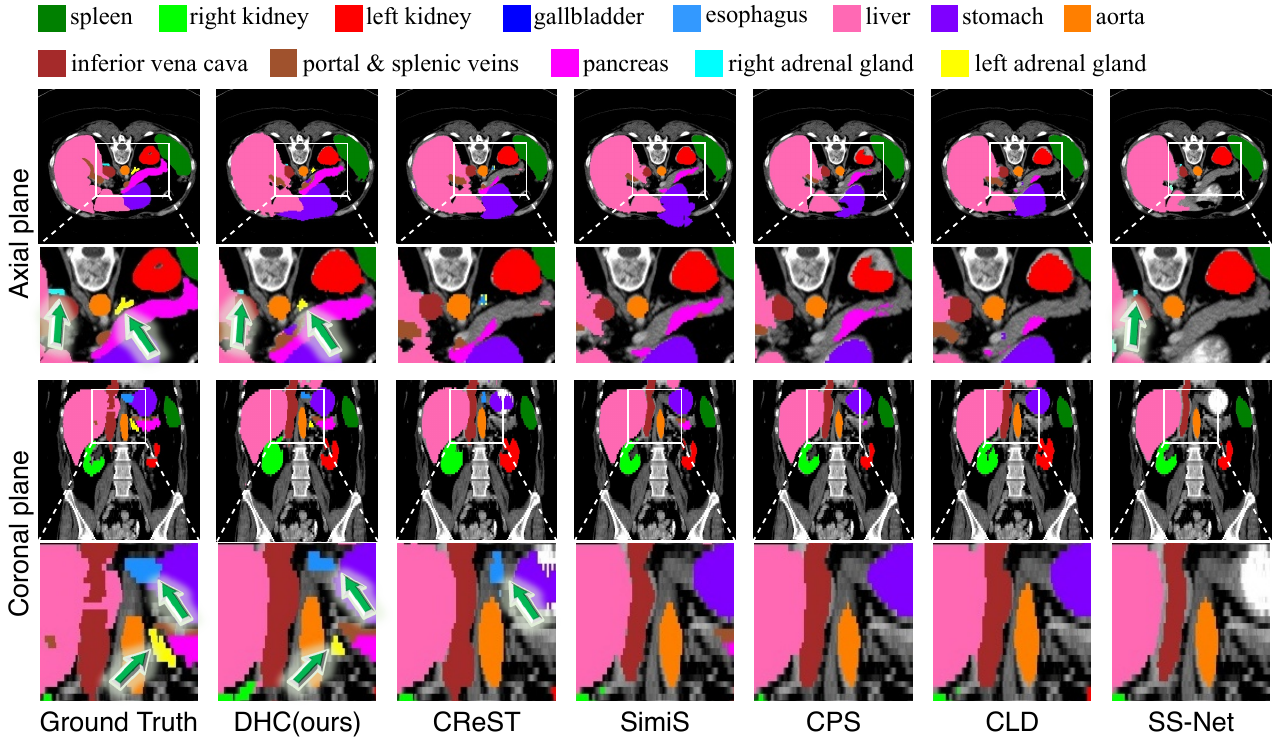} 
\caption{Qualitative comparison between DHC and the SOTA methods on \textbf{20\% labeled Synapse dataset}. The \textcolor[RGB]{82,179,94}{green}  arrows indicate the some minority classes being segmented.}
\label{vis_SOTA}
\end{figure}

\begin{table}[t]
\scriptsize
\caption{Quantitative comparison between DHC and SOTA SSL segmentation methods on \textbf{5\% labeled AMOS dataset}. 
}
\label{sota2}
\resizebox*{\linewidth}{!}{

\begin{tabular}{c|c|c@{\ \ }c|cccccccc@{\ }ccccccc}
\toprule
\multicolumn{2}{c|}{\multirow{2}{*}{Methods}}  & Avg. & Avg. & \multicolumn{15}{c}{Average Dice of Each   Class}                                        \\ 
\multicolumn{2}{c|}{}                          &    Dice                        &      ASD                     & Sp   & RK   & LK   & Ga   & Es   & Li   & St   & Ao   & IVC   & PA   & RAG  & LAG  & Du   & Bl  & P/U  \\\midrule
\multirow{9}{*}{\rotatebox{90}{General}}         & V-Net (fully)      & 76.50 	&2.01 	&92.2	&92.2	&93.3	&65.5	&70.3	&95.3	&82.4	&91.4	&85.0	&74.9	&58.6	&58.1	&65.6	&64.4	&58.3 \\ \midrule

& UA-MT~\cite{yu2019uamt}$^\dagger$      &42.16 	&15.48 	&59.8 	&64.9 	&64.0 	&35.3 	&\underline{34.1} 	&77.7 	&37.8 	&61.0 	&46.0 	&33.3 	&\underline{26.9} 	&12.3 	&18.1 	& 29.7	&31.6  \\

& URPC~\cite{luo2021urpc}$^\dagger$       & 44.93 	&27.44 	&67.0 	&64.2 	&67.2 	&36.1 	&\textcolor{red}{0.0} 	&83.1 	&\underline{45.5} 	&67.4 	&54.4 	&\textbf{46.7} 	&\textcolor{red}{0.0} 	&\textbf{29.4} 	&\textbf{35.2} 	&44.5	&33.2  \\

& CPS~\cite{chen2021cps}$^\dagger$        & 41.08 	&20.37 	&56.1 	&60.3 &	59.4 	&33.3 &	25.4 	&73.8 	&32.4 	&65.7 	&52.1 	&31.1 	&25.5 	&6.2 	&18.4 	&40.7	&35.8  \\

& SS-Net~\cite{wu2022ssnet}$^\dagger$    & 33.88 	&54.72 	&65.4 	&68.3 	&69.9 	&37.8 &\textcolor{red}{0.0} 	&75.1 	&33.2 	&68.0 	&\underline{56.6} 	&33.5 	&\textcolor{red}{0.0} 	&\textcolor{red}{0.0}	&\textcolor{red}{0.0} 	&0.2	&0.2  \\

& DST~\cite{chen2022dst}$^\star$        & 41.44 	&21.12 	&58.9 	&63.3 	&63.8 	&37.7 	&29.6 	&74.6 	&36.1 	&66.1 	&49.9 	&32.8 	&13.5 	&5.5 	&17.6 	&39.1	&33.1  \\

& DePL~\cite{wang2022depl}$^\star$       & 41.97 	&20.42 	&55.7 	&62.4 	&57.7 	&36.6 	&31.3 	&68.4 	&33.9 	&65.6 	&51.9 	&30.2 	&23.3 	&10.2 	&20.9 	&43.9	&\textbf{37.7}  \\ \midrule
\multirow{6}{*}{\rotatebox{90}{Imbalance}} 

& Adsh~\cite{guo2022adsh}$^\star$       & 40.33	& 24.53	& 56.0	& 63.6	& 57.3	& 34.7	& 25.7	& 73.9	& 30.7	& 65.7	& 51.9	& 27.1	& 20.2	& 0.0	& 18.6	& 43.5	& 35.9  \\ 

& CReST~\cite{wei2021crest}$^\star$      & 46.55	& 14.62	& 66.5	& 64.2	& 65.4	& 36.0	& 32.2	& 77.8	& 43.6	& 68.5	& 52.9	& 40.3	& 24.7	& 19.5	& 26.5	& 43.9	& 36.4  \\

& SimiS~\cite{simis}$^\star$      & \underline{47.27}	& \textbf{11.51}	& \textbf{77.4}	& \textbf{72.5}	& 68.7	& 32.1	& 14.7	& \textbf{86.6}	& \textbf{46.3}	& \textbf{74.6}	& 54.2	& 41.6	& 24.4	& 17.9	& 21.9	& \textbf{47.9}	& 28.2  \\

& Basak \textit{et al.}~\cite{basak2022addressing}$^\dagger$        & 38.73	& 31.76	& \underline{68.8}	& 59.0	& 54.2	& 29.0	& \textcolor{red}{0.0}	& \underline{83.7}	& 39.3	& 61.7	& 52.1	& 34.6	& \textcolor{red}{0.0}	& \textcolor{red}{0.0}	& 26.8	& \underline{45.7}	& 26.2   \\
                                 
& CLD~\cite{lin2022cld}$^\dagger$        & 46.10	& 15.86	& 67.2	& 68.5	& \textbf{71.4}	& \underline{41.0}	& 21.0	& 76.1	& 42.4	& 69.8	& 52.1	& 37.9	& 24.7	& 23.4	& 22.7	& 38.1	& 35.2  \\

& \textbf{DHC (ours)}   & \textbf{49.53}	& \underline{13.89}	& 68.1	& \underline{69.6}	& \underline{71.1}	& \textbf{42.3}	& \textbf{37.0}	& 76.8	& 43.8	& \underline{70.8}	& \textbf{57.4}	& \underline{43.2}	& \textbf{27.0}	& \underline{28.7}	& \underline{29.1}	& 41.4	& \underline{36.7} \\ \bottomrule
\end{tabular}
}
\begin{threeparttable}
 \begin{tablenotes}
        \scriptsize
        \item[$\dagger$] we implement semi-supervised segmentation methods on our dataset.
        \item[$\star$] we extend semi-supervised classification methods to segmentation with CPS as the baseline.
\end{tablenotes}
\end{threeparttable}

\end{table}

\begin{table}[t]
\scriptsize
\caption{Ablation study on \textbf{20\% labeled Synapse dataset}. 
The first four rows show the effectiveness of our proposed methods; the last four rows verify the importance of the heterogeneous design by combining our proposed modules with the existing class-imbalance methods, and serve as detailed results of the $3^{rd}$ and $4^{th}$ columns in Fig.~\ref{heterogeneous}.}
\label{ablation_v2}
\resizebox*{\linewidth}{!}{
\begin{tabular}{c|cc|cccccccc@{\ }ccccc}
\toprule
{\multirow{2}{*}{ModelA-ModelB}}  & \multirow{2}{*}{Avg. Dice} & \multirow{2}{*}{Avg. ASD} & \multicolumn{13}{c}{Average Dice of Each   Class}                                        \\ 
& & & Sp   & RK   & LK   & Ga   & Es   & Li   & St   & Ao   & IVC  & PSV  & Pa   & RAG  & LAG  \\\midrule

Org-Org (CPS)       & 33.55±3.65	&41.21±9.08	& 62.8	& 55.2	& 45.4	& 35.9	& 0.0	& \textbf{91.1}	& 31.3	& 41.9	& 49.2	& 8.8	& 14.5	& 0.0	& 0.0  \\

\textcolor[RGB]{153,97,132}{DistDW}-\textcolor[RGB]{153,97,132}{DistDW}        & 43.41±1.46	&\underline{17.39±1.93}	& 56.1	& 66.7	& \underline{60.2}	& 40.3	& \textbf{23.5}	& 75.7	& 10.3	& \underline{70.1}	& \underline{61.2}	& 30.4	& 26.4	& \textbf{32.2}	& \underline{11.2} \\

\textcolor[RGB]{33,101,168}{DiffDW}-\textcolor[RGB]{33,101,168}{DiffDW} & 42.75±0.6	&18.64±5.17	& \underline{71.9}	& 65.7	& 49.8	& 58.9	& 6.7	& 88.3	& 32.0	& 59.5	& 51.8	& 29.0	& 13.4	& 22.6	& 6.2 \\

\textbf{\textcolor[RGB]{33,101,168}{DiffDW}-\textcolor[RGB]{153,97,132}{DistDW} }  & \textbf{48.61±0.91}	&\textbf{10.71±2.62}	& 62.8	& 69.5	& 59.2	& 66.0	& \underline{13.2}	& 85.2	& \underline{36.9}	& 67.9	& \textbf{61.5}	& \textbf{37.0}	& \textbf{30.9}	& \underline{31.4}	& 10.6 \\ \midrule

\textcolor[RGB]{153,97,132}{DistDW}-CReST & 45.61±2.98	&21.58±3.89	& 68.7	& \textbf{70.5}	& 58.6	& 60.0	& 12.4	& 79.7	& 31.5	& 60.9	& 58.3	& 30.7	& \underline{29.0}	& 27.4	& 5.3  \\

\textcolor[RGB]{153,97,132}{DistDW}-SimiS & 41.92±2.52	&30.29±4.49	& 70.3	& 68.9	& \textbf{63.6}	& 56.2	& 3.1	& 77.0	& 14.0	& \textbf{75.4}	& 57.1	& 26.8	& 27.5	& 3.3	& 1.7  \\

\textcolor[RGB]{33,101,168}{DiffDW}-CReST & \underline{46.52±1.5}	&19.47±2.1	& 59.5	& 64.6	& 60.1	& \underline{67.3}	& 0.0	& 87.2	& 34.4	& 65.5	& 60.4	& 31.9	& 28.5	& 30.6	& \textbf{14.8} \\

\textcolor[RGB]{33,101,168}{DiffDW}-SimiS & 43.85±0.5	& 32.55±1.4	& \textbf{73.5}	& \underline{70.2}	& 54.9	& \textbf{69.7}	& 0.0	& \underline{89.4}	& \textbf{41.1}	& 67.5	& 51.8	& \underline{34.8}	& 16.7	& 0.6	& 0.0 \\
                                 
\bottomrule
\end{tabular}
}
\end{table}

\subsubsection{Comparison with State-of-the-Art Methods.}
We compare our method with several state-of-the-art semi-supervised segmentation methods~\cite{yu2019uamt,luo2021urpc,wu2022ssnet,chen2021cps}. 
Moreover, simply extending the state-of-the-art semi-supervised classification methods~\cite{wei2021crest,simis,guo2022adsh,chen2022dst,wang2022depl}, including class-imbalanced designs~\cite{wei2021crest,simis,guo2022adsh} to segmentation, is a straightforward solution to our task. Therefore, we extend these methods to segmentation with CPS as the baseline. 
As shown in Table~\ref{sota1}\&\ref{sota2}, the general semi-supervised methods which do not consider the class imbalance problem fail to capture effective features of the minority classes and lead to terrible performances (\textcolor{red}{colored with red}). 
The methods considered the class imbalance problem have better results on some smaller minority classes such as gallbladder, portal \& splenic veins and \textit{etc.}
However, they still fail in some minority classes (Es, RAG, and LAG) since this task is highly imbalanced. 
Our proposed DHC outperforms these methods, especially in those classes with very few samples.
Note that our method performs better than the fully-supervised method for the RAG segmentation. 
Furthermore, our method outperforms SOTA methods on Synapse by larger margins than the AMOS dataset, demonstrating the more prominent stability and effectiveness of the proposed DHC framework in scenarios with a severe lack of data.
Visualization results in Fig.~\ref{vis_SOTA} show our method performs better on minority classes which are pointed with green arrows.
More results on datasets with different labeled ratios can be found in the supplementary material.

\subsubsection{Ablation Study.}\label{ablation}
To validate the effectiveness of the proposed DHC framework and the two learning strategies, DistDW and DiffDW, we conduct ablation experiments, as shown in Table~\ref{ablation_v2}. 
DistDW (`\textcolor[RGB]{153,97,132}{DistDW}-\textcolor[RGB]{153,97,132}{DistDW}') alleviates the bias of baseline on majority classes and thus segments the minority classes (RA, LA, ES, \textit{etc.}) very well. However, it has unsatisfactory results on the spleen and stomach, which are difficult classes but down-weighted due to the larger voxel numbers.
DiffDW (`\textcolor[RGB]{33,101,168}{DiffDW}-\textcolor[RGB]{33,101,168}{DiffDW}') shows complementary results with DistDW, it has better results on difficult classes (\eg, stomach since it is hollow inside).
When combining these two weighting strategies in a heterogeneous co-training way (`\textcolor[RGB]{33,101,168}{DiffDW}-\textcolor[RGB]{153,97,132}{DistDW}', namely DHC), the Dice score has 5.12\%, 5.6\% and 13.78\% increase compared with DistDW, DiffDW, and the CPS baseline.
These results highlight the efficacy of incorporating heterogeneous information in avoiding over-fitting and enhancing the performance of the CPS baseline.

\section{Conclusion}
This work proposes a novel Dual-debiased Heterogeneous Co-training framework for class-imbalanced semi-supervised segmentation. We are the first to state the homogeneity issue of CPS and solve it intuitively in a heterogeneous way.
To achieve it, we propose two diverse and accurate weighting strategies: DistDW for eliminating the data bias of majority classes and DiffDW for eliminating the learning bias of well-performed classes.
By combining the complementary properties of DistDW and DiffDW, the overall framework can learn both the minority classes and the difficult classes well in a balanced way.
Extensive experiments show that the proposed framework brings significant improvements over the baseline and outperforms previous SSL methods considerably.

\section*{Acknowledgement}
This work was supported in part by a grant from Hong Kong Innovation and Technology Commission (Project no. ITS/030/21) and in part by a research grant from Beijing Institute of Collaborative Innovation (BICI) under collaboration with HKUST under Grant HCIC-004 and in part by grants from Foshan HKUST Projects under Grants FSUST21-HKUST10E and FSUST21-HKUST11E.

\bibliographystyle{splncs04}
\bibliography{miccai_bib}
%





\newpage

\title{Supplementary Material}

\author{Haonan Wang, Xiaomeng Li\textsuperscript{(\Letter)}} 

\institute{Department of Electronic and Computer Engineering, The Hong Kong University
of Science and Technology, Hong Kong, China
 \\ \email{eexmli@ust.hk}}

\maketitle

\section*{Training Details}
We implement the proposed framework with PyTorch, using a single NVIDIA A100 GPU. The network parameters are optimized with SGD with a momentum of 0.9 and an initial learning rate of 0.03. We employ a “poly” decay strategy follow~\cite{isensee2021nnunet}.
In the training stage, simple data augmentations (random cropping and random flipping) are used to avoid over-fitting.
We trained the networks 300 epochs with batch size 4, consisting of 2 labeled and 2 unlabeled data.
In the inference stage, we use the average of the two sub-networks for prediction for all the CPS-based methods to avoid over-fitting to one of the perturbations.
We evaluate the prediction of the network with two metrics, including Dice and the average surface distance (ASD).
Final segmentation results are obtained using a sliding window strategy with a stride size of $32 \times 32 \times 16$.

\begin{table}[!ht]
\scriptsize
\caption{Quantitative comparison between DHC and SOTA SSL segmentation methods on \textbf{10\% labeled Synapse dataset}. `General' or `Imbalance' indicate whether the methods consider class-imbalance issue or not.
Sp: spleen, RK: right kidney, LK: left kidney, Ga: gallbladder, Es: esophagus, Li: liver, St: stomach, Ao: aorta, IVC: inferior vena cava, PSV: portal \& splenic veins, Pa: pancreas, RAG: right adrenal gland, LAG: left adrenal gland.
}
\label{sota}
\resizebox*{\linewidth}{!}{
\begin{tabular}{c|c|cc|cccccccc@{\ }ccccc}
\toprule
\multicolumn{2}{c|}{\multirow{2}{*}{Methods}}  & \multirow{2}{*}{Avg. Dice} & \multirow{2}{*}{Avg. ASD} & \multicolumn{13}{c}{Average Dice of Each   Class}                                        \\ 
\multicolumn{2}{c|}{}                          &                            &                           & Sp   & RK   & LK   & Ga   & Es   & Li   & St   & Ao   & IVC  & PSV  & PA   & RAG  & LAG  \\\midrule
\multirow{9}{*}{\rotatebox{90}{General}}         & V-Net (fully)      & 62.09±1.2	&10.28±3.9	& 84.6	& 77.2	& 73.8	& 73.3	& 38.2	& 94.6	& 68.4	& 72.1	& 71.2	& 58.2	& 48.5	& 17.9	& 29.0 \\ \midrule

& UA-MT~\cite{yu2019uamt}$^\dagger$     &18.07±1.2	&57.64±1.8	& 27.1	& 7.1	& 17.0	& 24.4	& \textcolor{red}{0.0}	& 80.6	& 15.6	& 39.3	& 16.7	& 4.4	& 2.7	& \textcolor{red}{0.0}	& \textcolor{red}{0.0} \\

& URPC~\cite{luo2021urpc}$^\dagger$       &26.37±1.5	&53.95±11.3	& \textbf{51.7}	& 35.1	& 26.4	& 7.3	& \textcolor{red}{0.0}	& \textbf{83.8}	& 21.3	& \textbf{69.0}	& \textbf{41.0}	& 1.9	& 5.2	& \textcolor{red}{0.0} 	&\textcolor{red}{0.0}  \\

& CPS~\cite{chen2021cps}$^\dagger$        &21.96±1.2	&55.42±4.6	& 37.9	& 31.8	& 19.0	& 31.9	& \textcolor{red}{0.0}	& 65.1	& 15.5	& 44.8	& 29.6	& 4.3	& 5.5	& \textcolor{red}{0.0}	& \textcolor{red}{0.0}  \\

& SS-Net~\cite{wu2022ssnet}$^\dagger$     &17.5±3.0	&66.17±8.0	& 45.6	& 11.6	& \textbf{42.3}	& 2.4	& \textcolor{red}{0.0}	& 74.5	& 6.0	& 32.6	& 2.8	& \textcolor{red}{0.0}	& \textcolor{red}{0.0}	& 3.8	& 5.8  \\

& DST~\cite{chen2022dst}$^\star$        &20.91±5.9	&61.33±18.3	& 43.3	& 32.8	& 16.0	& 24.9	& \textcolor{red}{0.0}	& 75.8	& \textbf{22.4}	& 27.6	& 19.4	& 3.8	& 5.4	& 0.3	& \textcolor{red}{0.0}\\

& DePL~\cite{wang2022depl}$^\star$       &21.01±3.3	&58.42±6.3	& 34.2	& 32.2	& 17.3	& 27.2	& \textcolor{red}{0.0}	& 65.7	& 16.8	& 40.8	& 29.3	& 2.8	& 6.8	& \textcolor{red}{0.0}	& \textcolor{red}{0.0} \\ \midrule
\multirow{8}{*}{\rotatebox{90}{Imbalance}} 

& Adsh~\cite{guo2022adsh}$^\star$       &22.8±0.9	&46.18±4.0	& 36.0	& 35.7	& 20.0	& 31.0	& \textcolor{red}{0.0}	& 74.7	& 18.3	& 32.3	& 27.8	& 11.7	& 7.3	& 1.7	& \textcolor{red}{0.0}  \\ 

& CReST~\cite{wei2021crest}$^\star$      &26.56±2.9	&36.17±1.0	& 37.3	& 46.5	& 25.2	& 27.1	& 1.7	& 66.3	& 14.2	& 45.2	& 35.8	& 11.2	& 6.8	& 24.2	& 3.8  \\

& SimiS~\cite{simis}$^\star$      & 25.05±3.1	&43.93±2.4	& 42.0	& 38.6	& 27.2	& 19.7	& \textcolor{red}{0.0}	& 74.2	& 16.5	& 51.7	& 35.0	& 13.6	& 5.4	& \textcolor{red}{0.0}	& 1.8  \\

& Basak \textit{et al.}~\cite{basak2022addressing}$^\dagger$         &25.3±2.2	&50.02±5.7	& 40.9	& 42.3	& 19.2	& 35.2	& \textcolor{red}{0.0}	& 75.7	& 19.2	& 44.7	& 32.8	& 5.0	& \textbf{10.4}	& 3.5	& \textcolor{red}{0.0}  \\
                                 
& CLD~\cite{lin2022cld}$^\dagger$        & 22.49±1.6	&49.74±4.1	& 39.3	& 43.9	& 25.6	& 12.8	& \textcolor{red}{0.0}	& 73.3	& 14.3	& 41.1	& 25.7	& 8.8	& 6.1	& 0.2	& 1.1  \\

& \textbf{DistDW(ours)}   & 27.21±0.9	&32.38±4.0	& 47.8	& 42.3	& 33.1	& 27.0	& 1.1	& 65.5	& 20.7	& 49.0	& 34.5	& 8.1	& 7.4	& 14.5	& 2.8 \\

& \textbf{DiffDW(ours)}   & 28.63±2.5	&24.81±4.0	& 44.1	& 33.4	& 25.3	& \textbf{37.0}	& 6.3	& 75.8	& 19.1	& 46.3	& 28.6	& \textbf{17.5}	& 7.8	& \textbf{24.7}	& 6.3\\
                                  
& \textbf{DHC(ours)}   & \textbf{31.64±0.9}	&\textbf{21.82±1.0}	& 45.1	& \textbf{47.4}	& 33.1	& 36.6	& \textbf{7.1}	& 71.4	& 17.8	& 58.9	& 34.4	& 16.5	& 9.3	& 21.8	& \textbf{12.0} \\ \bottomrule
\end{tabular}
}
\begin{threeparttable}
 \begin{tablenotes}
        \scriptsize
        \item[$\dagger$] we implement semi-supervised segmentation methods on our dataset.
        \item[$\star$] we extend semi-supervised classification methods to segmentation with CPS as the baseline.
\end{tablenotes}
\end{threeparttable}
\end{table}

\begin{table}[!ht]
\scriptsize
\caption{Quantitative comparison between DHC and SOTA SSL segmentation methods on\textbf{ 40\% labeled Synapse dataset}. 
}
\label{sota}
\resizebox*{\linewidth}{!}{
\begin{tabular}{c|c|cc|cccccccc@{\ }ccccc}
\toprule
\multicolumn{2}{c|}{\multirow{2}{*}{Methods}}  & \multirow{2}{*}{Avg. Dice} & \multirow{2}{*}{Avg. ASD} & \multicolumn{13}{c}{Average Dice of Each   Class}                                        \\ 
\multicolumn{2}{c|}{}                          &                            &                           & Sp   & RK   & LK   & Ga   & Es   & Li   & St   & Ao   & IVC  & PSV  & PA   & RAG  & LAG  \\\midrule
\multirow{9}{*}{\rotatebox{90}{General}}         & V-Net (fully)      & 62.09±1.2	&10.28±3.9	& 84.6	& 77.2	& 73.8	& 73.3	& 38.2	& 94.6	& 68.4	& 72.1	& 71.2	& 58.2	& 48.5	& 17.9	& 29.0 \\ \midrule

& UA-MT~\cite{yu2019uamt}$^\dagger$      &17.09±2.97	&91.86±7.93	& 4.2	& 57.8	& 32.2	& \textcolor{red}{0.0}	& \textcolor{red}{0.0}	& 91.0	& 37.0	& \textcolor{red}{0.0}	& \textcolor{red}{0.0}	& \textcolor{red}{0.0}	& \textcolor{red}{0.0}	& \textcolor{red}{0.0}	& \textcolor{red}{0.0}  \\

& URPC~\cite{luo2021urpc}$^\dagger$       &24.83±8.19	&74.44±17.01	& 42.6	& 44.8	& 51.4	& \textcolor{red}{0.0}	& \textcolor{red}{0.0}	& 86.7	& 37.8	& 25.8	& 27.0	& \textcolor{red}{0.0}	& 6.6	& \textcolor{red}{0.0}	& \textcolor{red}{0.0}  \\

& CPS~\cite{chen2021cps}$^\dagger$       &33.07±1.07	&60.46±2.25	& 68.4	& 72.7	& 64.2	& \textcolor{red}{0.0}	& \textcolor{red}{0.0}	& 91.9	& 42.1	& 66.2	& 22.3	& \textcolor{red}{0.0}	& 2.2	& \textcolor{red}{0.0}	& \textcolor{red}{0.0} \\

& SS-Net~\cite{wu2022ssnet}$^\dagger$     &32.98±10.99	&71.18±20.77	& 49.0	& 68.9	& 71.4	& 22.9	& \textcolor{red}{0.0}	& 92.0	& 34.7	& 51.7	& 38.1	& \textcolor{red}{0.0}	& \textcolor{red}{0.0}	& \textcolor{red}{0.0}	& \textcolor{red}{0.0} \\

& DST~\cite{chen2022dst}$^\star$        &35.57±1.54	&55.69±1.43	& 73.8	& \textbf{73.2}	& 64.2	& \textcolor{red}{0.0}	& \textcolor{red}{0.0}	& \textbf{92.1}	& 41.3	& 71.8	& 40.7	& \textcolor{red}{0.0}	& 5.2	& \textcolor{red}{0.0}	& \textcolor{red}{0.0}  \\

& DePL~\cite{wang2022depl}$^\star$       &36.16±2.08	&56.14±7.61	& 72.7	& 72.4	& 64.4	& 13.3	& \textcolor{red}{0.0}	& 91.7	& 42.8	& 63.8	& 47.0	& \textcolor{red}{0.0}	& 1.9	& \textcolor{red}{0.0}	& \textcolor{red}{0.0}  \\ \midrule
\multirow{8}{*}{\rotatebox{90}{Imbalance}} 

& Adsh~\cite{guo2022adsh}$^\star$       &35.91±6.17	&53.7±6.95	& 66.8	& 72.5	& 64.4	& 19.1	& \textcolor{red}{0.0}	& 91.8	& 43.8	& 62.0	& 39.8	& \textcolor{red}{0.0}	& 6.5	& \textcolor{red}{0.0}	& \textcolor{red}{0.0}  \\ 

& CReST~\cite{wei2021crest}$^\star$      &41.6±2.49	&27.82±5.07	& 53.8	& 69.5	& 58.1	& 35.3	& 17.7	& 85.0	& 36.0	& 60.3	& 45.2	& 21.5	& 24.2	& 23.3	& 10.8 \\

& SimiS~\cite{simis}$^\star$      &47.09±2.33	&33.46±1.75	& 75.4	& 66.9	& 69.0	& 62.6	& \textcolor{red}{0.0}	& 81.6	& \textbf{53.1}	& 80.4	& 56.2	& 29.9	& 37.1	& \textcolor{red}{0.0}	& \textcolor{red}{0.0} \\

& Basak \textit{et al.}~\cite{basak2022addressing}$^\dagger$   &35.03±3.68	&60.69±6.57	& 69.1	& 72.8	& 67.5	& \textcolor{red}{0.0}	& \textcolor{red}{0.0}	& 91.6	& 45.0	& 62.2	& 39.1	& \textcolor{red}{0.0}	& 8.0	& \textcolor{red}{0.0}	& \textcolor{red}{0.0}   \\
                                 
& CLD~\cite{lin2022cld}$^\dagger$        &48.23±1.02	&28.79±3.64	& 78.3	& 73.1	& \textbf{73.8}	& 57.0	& \textcolor{red}{0.0}	& 87.6	& 51.7	& 82.8	& 50.5	& 38.0	& 31.8	& 1.5	& 0.8 \\

& \textbf{DistDW(ours)}   &51.86±4.13	&14.68±4.31	& 78.0	& 69.7	& 66.2	& 65.9	& \textbf{34.4}	& 74.3	& 23.2	& 76.9	& 58.0	& 32.4	& 33.4	& 29.6	& 32.1 \\

& \textbf{DiffDW(ours)}   &50.84±2.78	&19.24±11.48	& 76.7	& 67.8	& 68.4	& 58.6	& 3.2	& 82.5	& 42.8	& 83.6	& 51.0	& 44.2	& 42.3	& 22.9	& 16.9 \\
                                
& \textbf{DHC(ours)}   &\textbf{57.13±0.8}	&\textbf{11.66±2.7}	& \textbf{82.5}	& 72.8	& 73.5	& \textbf{69.8}	& 10.7	& 71.9	& 41.2	& \textbf{83.7}	& \textbf{66.1}	& \textbf{53.8}	& \textbf{47.4}	& \textbf{36.8}	& \textbf{32.7} \\ \bottomrule
\end{tabular}
}
\begin{threeparttable}
 \begin{tablenotes}
        \scriptsize
        \item[$\dagger$] we implement semi-supervised segmentation methods on our dataset.
        \item[$\star$] we extend semi-supervised classification methods to segmentation with CPS as the baseline.
\end{tablenotes}
\end{threeparttable}
\end{table}

\begin{table}[!ht]
\scriptsize
\caption{Quantitative comparison between DHC and SOTA SSL segmentation methods on \textbf{2\% labeled AMOS dataset}. 
}
\label{sota2}
\resizebox*{\linewidth}{!}{

\begin{tabular}{c|c|c@{\ \ }c|cccccccc@{\ }ccccccc}
\toprule
\multicolumn{2}{c|}{\multirow{2}{*}{Methods}}  & Avg. & Avg. & \multicolumn{15}{c}{Average Dice of Each   Class}                                        \\ 
\multicolumn{2}{c|}{}                          &    Dice                        &      ASD                     & Sp   & RK   & LK   & Ga   & Es   & Li   & St   & Ao   & IVC   & PA   & RAG  & LAG  & Du   & Bl  & P/U  \\\midrule
\multirow{9}{*}{\rotatebox{90}{General}}         & V-Net (fully)      & 76.50 	&2.01 	&92.2	&92.2	&93.3	&65.5	&70.3	&95.3	&82.4	&91.4	&85.0	&74.9	&58.6	&58.1	&65.6	&64.4	&58.3 \\ \midrule

& UA-MT~\cite{yu2019uamt}$^\dagger$      & 33.96	& 22.43	& \textbf{62.5}	& \textbf{61.7}	& \textbf{59.8}	& 17.5	& 13.8	& 73.4	& 39.4	& 34.6	& 32.4	& 26.5	& 12.1	& 6.5	& 15.3	& 32.4	& 21.7  \\

& URPC~\cite{luo2021urpc}$^\dagger$       & \textbf{38.39}	& 37.58	& 60.8	& 57.7	& 56.5	& \textbf{34.6}	& \textcolor{red}{0.0}	& \textbf{78.4}	& \textbf{41.4}	& \textbf{53.3}	& \textbf{49.6}	& \textbf{40.4}	& \textcolor{red}{0.0}	& \textcolor{red}{0.0}	& \textbf{30.1}	& 42.5	& \textbf{30.6}  \\

& CPS~\cite{chen2021cps}$^\dagger$        & 31.78	& 39.23	& 55.9	& 46.9	& 53.1	& 27.7	& \textcolor{red}{0.0}	& 66.4	& 25.2	& 41.8	& 45.2	& 29.4	& 0.1	& \textcolor{red}{0.0}	& 22.1	& 38.7	& 24.2  \\

& SS-Net~\cite{wu2022ssnet}$^\dagger$    & 17.47	& 59.05	& 37.7	& 20.1	& 26.3	& 9.0	& 3.3	& 57.1	& 25.1	& 28.4	& 28.2	& \textcolor{red}{0.0}	& \textcolor{red}{0.0}	& \textcolor{red}{0.0}	& \textcolor{red}{0.0}	& 26.5	& 0.2  \\

& DST~\cite{chen2022dst}$^\star$        & 31.94	& 39.15	& 50.9	& 52.4	& 56.9	& 24.6	& \textcolor{red}{0.0}	& 59.4	& 31.5	& 41.8	& 43.1	& 26.2	& \textcolor{red}{0.0}	& \textcolor{red}{0.0}	& 23.8	& 42.6	& 25.9  \\

& DePL~\cite{wang2022depl}$^\star$       & 31.56	& 40.70	& 57.1	& 49.3	& 54.3	& 26.6	& 0.1	& 69.2	& 26.2	& 41.1	& 46.7	& 23.9	& \textcolor{red}{0.0}	& \textcolor{red}{0.0}	& 16.7	& 40.3	& 21.8  \\ \midrule
\multirow{6}{*}{\rotatebox{90}{Imbalance}} 

& Adsh~\cite{guo2022adsh}$^\star$       & 30.30	& 42.48	& 53.9	& 45.1	& 51.2	& 28.5	& \textcolor{red}{0.0}	& 62.1	& 27.0	& 41.4	& 42.7	& 25.0	& \textcolor{red}{0.0}	& \textcolor{red}{0.0}	& 20.3	& 35.8	& 21.6  \\ 

& CReST~\cite{wei2021crest}$^\star$      & 34.13	& 20.15	& 57.9	& 51.5	& 49.1	& 22.7	& 13.2	& 66.2	& 34.4	& 39.4	& 40.4	& 24.6	& 17.2	& 10.2	& 24.4	& 36.5	& 24.4  \\

& SimiS~\cite{simis}$^\star$     & 36.89	& 26.16	& 57.8	& 58.6	& 58.6	& 22.9	& \textcolor{red}{0.0}	& 70.9	& 38.0	& 52.0	& 47.0	& 32.4	& 20.2	& \textbf{11.5}	& 18.1	& 39.9	& 25.5  \\

& Basak \textit{et al.}~\cite{basak2022addressing}$^\dagger$        & 29.87	& 35.55	& 50.7	& 47.7	& 44.1	& 21.1	& \textcolor{red}{0.0}	& 61.8	& 27.7	& 38.1	& 40.4	& 21.8	& 9.6	& 9.5	& 14.6	& 36.5	& 24.5   \\
                                 
& CLD~\cite{lin2022cld}$^\dagger$        & 36.23	& 27.63	& 55.8	& 55.8	& 59.1	& 23.9	& \textcolor{red}{0.0}	& 69.9	& 38.2	& 50.1	& 44.5	& 32.3	& 18.9	& 9.2	& 18.8	& 42.2	& 24.9  \\

& \textbf{DHC (ours)}   & 38.28	& \textbf{20.34}	& 62.1	& 59.5	& 57.8	& 25.0	& \textbf{20.5}	& 66.0	& 38.2	& 51.3	& 47.9	& 26.8	& \textbf{26.4}	& 7.0	& 17.8	& \textbf{43.2}	& 24.8 \\ \bottomrule
\end{tabular}
}
\begin{threeparttable}
 \begin{tablenotes}
        \scriptsize
        \item[$\dagger$] we implement semi-supervised segmentation methods on our dataset.
        \item[$\star$] we extend semi-supervised classification methods to segmentation with CPS as the baseline.
\end{tablenotes}
\end{threeparttable}

\end{table}

\begin{table}[!ht]
\scriptsize
\caption{Quantitative comparison between DHC and SOTA SSL segmentation methods on \textbf{10\% labeled AMOS dataset}. 
}
\label{sota2}
\resizebox*{\linewidth}{!}{

\begin{tabular}{c|c|c@{\ \ }c|cccccccc@{\ }ccccccc}
\toprule
\multicolumn{2}{c|}{\multirow{2}{*}{Methods}}  & Avg. & Avg. & \multicolumn{15}{c}{Average Dice of Each   Class}                                        \\ 
\multicolumn{2}{c|}{}                          &    Dice                        &      ASD                     & Sp   & RK   & LK   & Ga   & Es   & Li   & St   & Ao   & IVC   & PA   & RAG  & LAG  & Du   & Bl  & P/U  \\\midrule
\multirow{9}{*}{\rotatebox{90}{General}}         & V-Net (fully)      & 76.50 	&2.01 	&92.2	&92.2	&93.3	&65.5	&70.3	&95.3	&82.4	&91.4	&85.0	&74.9	&58.6	&58.1	&65.6	&64.4	&58.3 \\ \midrule

& UA-MT~\cite{yu2019uamt}$^\dagger$      & 40.60	& 38.45	& 61.0	& 75.4	& 58.8	& 0.1	& \textcolor{red}{0.0}	& 84.4	& 45.2	& 72.8	& 61.6	& 36.2	& \textcolor{red}{0.0}	& \textcolor{red}{0.0}	& 30.7	& 46.5	& 36.3  \\

& URPC~\cite{luo2021urpc}$^\dagger$       & 49.09	& 29.69	& 81.7	& 77.5	& 77.2	& 38.1	& \textcolor{red}{0.0}	& 87.7	& 57.9	& 75.0	& 62.7	& 52.1	& \textcolor{red}{0.0}	& \textcolor{red}{0.0}	& 35.9	& 48.8	& 41.7  \\

& CPS~\cite{chen2021cps}$^\dagger$        & 54.51	& 7.84	& 80.7	& 79.8	& 74.3	& 35.2	& 44.4	& 90.5	& 51.1	& 76.1	& 65.6	& 48.6	& 31.6	& 21.8	& 33.6	& 47.0	& 37.3  \\

& SS-Net~\cite{wu2022ssnet}$^\dagger$    & 38.91	& 53.43	& 73.4	& 73.4	& 72.2	& 42.4	& \textcolor{red}{0.0}	& 83.5	& 46.7	& 74.1	& 69.6	& \textcolor{red}{0.0}	& \textcolor{red}{0.0}	& \textcolor{red}{0.0}	&\textcolor{red}{0.0}	& 48.3	& 0.2 \\

& DST~\cite{chen2022dst}$^\star$        & 52.24	& 17.66	& 81.7	& 80.2	& 78.6	& 39.5	& 41.0	& 89.8	& 52.8	& 78.5	& 65.9	& 51.1	& 4.3	& 0.1	& 34.2	& 48.8	& 37.2  \\

& DePL~\cite{wang2022depl}$^\star$      & 56.76	& 6.70	& 81.9	& 80.6	& 79.5	& 41.0	& 42.6	& 89.3	& 57.6	& 79.1	& 66.0	& 53.2	& 34.6	& 21.8	& 34.9	& 48.4	& 40.9  \\ \midrule
\multirow{6}{*}{\rotatebox{90}{Imbalance}} 

& Adsh~\cite{guo2022adsh}$^\star$       & 54.92	& 8.07	& 81.6	& 78.5	& 76.6	& 40.1	& 43.4	& 90.1	& 53.0	& 76.7	& 64.4	& 48.3	& 25.9	& 24.2	& 34.7	& 48.7	& 37.7  \\ 

& CReST~\cite{wei2021crest}$^\star$      & 60.74	& 4.65	& 85.3	& 84.5	& 84.0	& 43.2	& 50.8	& 89.9	& 58.7	& 84.7	& 73.0	& 54.2	& \textbf{41.8}	& 31.6	& 41.0	& 52.8	& 35.8  \\

& SimiS~\cite{simis}$^\star$      & 57.48	& 4.46	& 83.1	& 80.9	& 80.0	& 39.6	& 45.9	& 90.0	& 57.1	& 78.0	& 66.3	& 54.1	& 35.8	& 26.9	& 39.9	& 49.3	& 35.4  \\

& Basak \textit{et al.}~\cite{basak2022addressing}$^\dagger$       & 53.66	& 8.50	& 80.3	& 78.2	& 79.0	& 36.3	& 40.3	& 88.6	& 53.2	& 76.8	& 65.6	& 46.8	& 23.9	& 16.1	& 31.4	& 49.7	& 38.6   \\
                                 
& CLD~\cite{lin2022cld}$^\dagger$        & 61.55	& 4.21	& 86.0	& 85.3	& 84.8	& 44.5	& 51.9	& \textbf{90.8}	& 59.7	& 83.7	& 73.1	& 55.7	& 40.2	& \textbf{37.2}	& 41.4	& 53.0	& 36.1 \\

& \textbf{DHC (ours)}   & \textbf{64.16}	& \textbf{3.51}	& \textbf{87.4}	& \textbf{86.6}	& \textbf{87.1}	& \textbf{45.8}	& \textbf{57.0}	& 89.8	& \textbf{64.7}	& \textbf{86.0}	& \textbf{75.0}	& \textbf{62.5}	& 39.8	& 36.8	& \textbf{44.0}	& \textbf{56.5}	& \textbf{43.6} \\ \bottomrule
\end{tabular}
}
\begin{threeparttable}
 \begin{tablenotes}
        \scriptsize
        \item[$\dagger$] we implement semi-supervised segmentation methods on our dataset.
        \item[$\star$] we extend semi-supervised classification methods to segmentation with CPS as the baseline.
\end{tablenotes}
\end{threeparttable}

\end{table}

\end{document}